\documentclass[12pt]{article}
\usepackage{hyperref}
\usepackage[english]{babel}
\usepackage{amsmath}
\usepackage{latexsym}
\usepackage{amssymb}
\usepackage[all]{xy}
\usepackage[dvips]{graphicx}
\usepackage{epsfig}
\usepackage{psfrag}

\numberwithin{equation}{section} \numberwithin{table}{section}
\numberwithin{figure}{section}

\begin{document}


\begin{titlepage}
  \begin{flushright}
  {\small CQUeST-2009-0274}
  \end{flushright}

  \begin{center}

    \vspace{20mm}

    {\LARGE \bf $R^{2}$ Corrections to Asymptotically Lifshitz Spacetimes}

    \vspace{10mm}

    Da-Wei Pang$^{\dag}$

    \vspace{5mm}
    {\small \sl $\dag$ Center for Quantum Spacetime, Sogang University}\\
    {\small \sl Seoul 121-742, Korea\\}
    {\small \tt pangdw`at'sogang.ac.kr}
    \vspace{10mm}
  \end{center}

\begin{abstract}
\baselineskip=18pt We study $R^{2}$ corrections to five-dimensional
asymptotically Lifshitz spacetimes by adding Gauss-Bonnet terms in
the effective action. For the zero-temperature backgrounds we obtain
exact solutions in both pure Gauss-Bonnet gravity and Gauss-Bonnet
gravity with non-trivial matter. The dynamical exponent undergoes
finite renormalization in the latter case. For the
finite-temperature backgrounds we obtain black brane solutions
perturbatively and calculate the ratio of shear viscosity to entropy
density $\eta/s$. The KSS bound is still violated but unlike the
relativistic counterparts, the causality of the boundary field
theory cannot be taken as a constraint.

\end{abstract}
\setcounter{page}{0}
\end{titlepage}

\pagestyle{plain} \baselineskip=19pt

\tableofcontents

\section{Introduction}
The AdS/CFT correspondence~\cite{Maldacena:1997re, Gubser:1998bc,
Witten:1998qj} relates conformal field theories to gravitational
dynamics in asymptotically AdS backgrounds. As a strong-weak
duality, it has yielded many important insights into the dynamics of
strongly coupled field theories. For instance, the hydrodynamic
behavior of finite-temperature field theory can be reflected in the
dual gravity side~\cite{Son:2007vk}. Recently, there has been a
great deal of progress in applying the AdS/CFT correspondence to
study condensed matter systems near a critical point, for reviews
see~\cite{Hartnoll:2009sz}.

There are many scale-invariant field theories in which time and
space can scale differently,
\begin{equation}
t~\rightarrow~\lambda^{z}t,~~~~~x^{i}~\rightarrow~\lambda x^{i},
\end{equation}
where $z$ is called the `dynamical exponent'. Such field theories
with anisotropic scaling symmetry are of interest in studying
condensed matter systems near a critical point. The corresponding
gravity duals have been investigated extensively and there are
mainly two concrete cases of interest till now. One is the theory
with Galilean boost symmetry as well as anisotropic scaling
symmetry, whose symmetry group is the Schr\"{o}dinger group for
$z=2$.  The gravity duals were obtained in~\cite{Son:2008ye,
Balasubramanian:2008dm} with the following metric
\begin{equation}
ds^{2}=-r^{4}(dx^{+})^2-2r^{2}dx^{+}dx^{-}+\frac{dr^{2}}{r^{2}}+r^{2}d\vec{x}^{2}.
\end{equation}
The embedding of this spacetime into string theory and the
finite-temperature generalizations have been successfully realized
in~\cite{Herzog:2008wg, Maldacena:2008wh, Adams:2008wt,
Hartnoll:2008rs}. Such backgrounds, both the zero-temperature case
and the finite-temperature case, can be obtained by performing the
``Null Melvin Twist''~\cite{Gimon:2003xk} on the corresponding
(black) D-brane configurations.

The other one has no boost symmetry, which is known as the Lifshitz
case, and the gravity duals were obtained in~\cite{Kachru:2008yh}
\begin{equation}
ds^{2}=L^{2}(-r^{2z}dt^{2}+\frac{dr^{2}}{r^{2}}+r^{2}d\vec{x}^2).
\end{equation}
Some other gravity solutions with similar anisotropic scale
invariance were studied in~\cite{SSPal}. Unlike the Schr\"{o}dinger
case, it is quite difficult to embed the Lifshitz background into
string theory, so is to find the finite temperature generalizations.
The embedding of Lifshitz-like fixed points into type IIB string
theory was discussed extensively in~\cite{Azeyanagi:2009pr}, based
on the D3-D7 solutions introduced in~\cite{Fujita:2009kw}. Since the
dilaton in the solution is not constant, the anisotropic scale
invariance only holds at the leading order of interactions. Some
no-go theorems for string duals of non-relativistic Lifshitz-like
theories were proposed quite recently in~\cite{Li:2009pf}, where the
authors argued that such gravity duals in the supergravities were
not possible. Black hole in asymptotically Lifshitz spacetimes were
discussed in~\cite{Danielsson:2009gi, Mann:2009yx, Bertoldi:2009vn,
Bertoldi:2009dt}, where most of the solutions were obtained
numerically and exact solutions could be found only in certain
specific examples.

Several aspects of non-relativistic holography were studied
in~\cite{Taylor:2008tg}, where it was observed that the Lifshitz
geometry was a solution of a gravity theory coupled with a massive
vector. Furthermore, it was found that the following action
\begin{equation}
S=\frac{1}{16\pi G_{d+2}}\int
d^{d+2}x\sqrt{-g}[R-2\Lambda-\frac{1}{2}\partial_{\mu}\phi\partial^{\mu}\phi
-\frac{1}{4}e^{\lambda\phi}F_{\mu\nu}F^{\mu\nu}]
\end{equation}
admitted a solution with anisotropic scaling symmetry
\begin{eqnarray}
&
&ds^{2}=L^{2}(-r^{2z}dt^{2}+\frac{dr^{2}}{r^{2}}+r^{2}\sum\limits^{d}_{i=1}dx^{2}_{i}),\nonumber\\
&
&F_{rt}=qe^{-\lambda\phi}r^{z-d-1},~~~e^{\lambda\phi}=r^{\lambda\sqrt{2(z-1)d}},\nonumber\\
& &\lambda^{2}=\frac{2d}{z-1},~~~q^{2}=2L^{2}(z-1)(z+d),\nonumber\\
& &\Lambda=-\frac{(z+d-1)(z+d)}{2L^{2}}.
\end{eqnarray}
The metric is Lifshitz-like but the dilaton is not constant, so it
cannot be seen as a genuine gravity dual of Lifshitz-fixed points.
However, such a solution is worth investigating, as it also
possesses exact solutions with finite temperature. Some properties
of the corresponding black branes were discussed
in~\cite{Pang:2009ad}.

In this paper we will study $R^{2}$ corrections to the above
mentioned Lifshitz-like gravity backgrounds. The $1/N$ effects in
non-relativistic gauge-gravity duality was investigated extensively
in~\cite{Adams:2008zk}, where they argued that the dynamical
exponent received finite renormalization and the ratio of shear
viscosity to entropy density weakly violated the celebrated KSS
bound $1/4\pi$. Here we first study the $R^{2}$ corrections to the
zero-temperature cases by solving the equations of motion
explicitly. We find that an exact solution of Lifshitz background
can be found in pure Gauss-Bonnet gravity, while the dynamical
exponent undergoes a finite renormalization when non-trivial matter
fields are included. We also obtain the corresponding black brane
solutions by perturbative methods and calculate the ratio of shear
viscosity to entropy density. The result also violates the KSS bound
and it reduces to the known result when the dynamical exponent
$z=1$. However, since the boundary field theory is non-relativistic,
the causality of the field theory cannot be taken as a constraint on
the ratio.

The rest of the paper is organized as follows: In Section 2 we
review some necessary backgrounds, including the renormalization of
the dynamical exponent and the solutions in asymptotically Lifshitz
spacetimes. The $R^{2}$ corrections to the zero-temperature case are
studied in Section 3 and the corrections to the black brane case are
studied in Section 4. We calculate the ratio of shear viscosity to
entropy density in Section 5 through the effective coupling of the
transverse gravitons in the dual gravity side. A summary and
discussion will be given in the final section.

\section{Some backgrounds}
In this section we will review some backgrounds which are necessary
for further investigations. One of the main results
in~\cite{Adams:2008zk}, that is, both the radius of curvature and
the dynamical exponent $z$ may be renormalized for non-relativistic
Lifshitz metric, will be summarized in Section~\ref{sec:2.1}. The
zero-temperature and black brane solutions in asymptotically
Lifshitz spacetimes, obtained in~\cite{Taylor:2008tg}, will be
reviewed in Section~\ref{sec:2.2}.

\subsection{Solutions with Lifshitz symmetry and invariant two-forms}\label{sec:2.1}
The renormalization of the curvature radius and the dynamical
exponent for both Schr\"{o}dinger and Lifshitz metrics was
demonstrated in~\cite{Adams:2008zk} in a beautiful way. Here we just
focus on the Lifshitz case, whose symmetry algebra contains the
following generators: Hamiltonian $H$, linear momenta $P_{i}$,
angular momenta $M_{ij}$ and a dilaton operator $D$. The commutators
among $H$, $P_{i}$ and $M_{ij}$ behave the same as usual, and the
dilaton operator $D$ has the following non-trivial commutators
\begin{equation}
[D,P_{i}]=iP_{i},~~~[D,H]=izH.
\end{equation}
We rewrite the $(d+2)-$dimensional gravity duals of Lifshitz fixed
points in~\cite{Kachru:2008yh} as
\begin{equation}
ds^{2}=L^{2}(-\frac{dt^{2}}{r^{2z}}+\frac{dr^{2}+d\vec{x}^2}{r^{2}}).
\end{equation}
The corresponding Killing vectors are
\begin{equation}
\label{2eq3} H=-i\partial_{t},~~~P_{i}=-i\partial_{i},
~~~M_{ij}=-i(x^{i}\partial_{j}-x^{j}\partial_{i}),~~~D=-i(zt\partial_{t}+x^{i}\partial_{i}+r\partial_{r}).
\end{equation}

Since we are trying to find Lifshitz-like solutions to the equations
of motion of gravity coupled to some matter sector, the Einstein
tensor $G_{\mu\nu}$ and the stress tensor $T_{\mu\nu}$ are symmetric
two-tensors invariant under the Lifshitz symmetries~(\ref{2eq3}).
The search for solutions may be simplified if we expand the Einstein
equations in a basis of such symmetric invariant two-forms. Let
$\tau=\tau_{\mu\nu}dx^{\mu}dx^{\nu}$ be a symmetric two-tensor
invariant under the Lifshitz symmetries, i.e.
$\mathcal{L}_{v}\tau_{\mu\nu}=0$ for all the Killing vectors $v$
in~(\ref{2eq3}). The symmetry of the two-tensor plus the
conservation of the stress tensor $\nabla^{\mu}\tau_{\mu\nu}=0$
imply
\begin{equation}
\label{2eq4}
\tau=\alpha\frac{dt^{2}}{r^{2z}}+\beta\frac{d\vec{x}^2}{r^{2}}
+(\frac{(d-2)\beta-z\alpha}{d-2+z})\frac{dr^{2}}{r^{2}},
\end{equation}
where $\alpha, \beta$ are two constants. Thus it is a two-parameter
family of conserved stress tensors.

Consider the action of gravity coupled to an arbitrary matter sector
\begin{equation}
S=\frac{1}{16\pi
G}\int\sqrt{-g}(R-2\Lambda)+S_{m}(g_{\mu\nu},\phi_{i}),
\end{equation}
where $S_{m}$ denotes the matter part of the action and $\phi_{i}$
stand for the matter fields. $S_{m}$ can also contain higher
derivative corrections to the Einstein-Hilbert action. The equations
of motion are
\begin{equation}
R_{\mu\nu}-\frac{1}{2}Rg_{\mu\nu}+\Lambda g_{\mu\nu}=-8\pi
GT_{\mu\nu},
\end{equation}
where $T_{\mu\nu}=\delta S_{m}/\delta g^{\mu\nu}$ include the usual
matter stress tensor and the contributions from the higher curvature
terms. The invariance of the stress tensor $T_{\mu\nu}$ sets a
non-trivial constraint on $\phi_{i}$. The simplest constraint is to
require that the fields themselves are invariant
$\mathcal{L}_{v}\phi_{i}=0$. However, this is not strictly necessary
if we want $T_{\mu\nu}$ to be invariant. In~\cite{Adams:2008zk}, it
was demanded that any gauge invariant observables must be invariant
under the full symmetry. In the next subsection we will see that
such a requirement can also be released.

Now let us focus on the Einstein equations. The left hand side is
automatically a conserved two-tensor invariant under the Lifshitz
symmetries. Thus it can be written in the form~(\ref{2eq4}). It can
be seen that $\alpha$ and $\beta$ are simple functions of $z$ and
$L$, whose explicit expressions will not be shown. On the other
hand, the stress tensor takes the following form
\begin{equation}
T_{\mu\nu}=\alpha(z,L,\phi_{i})\frac{dt^{2}}{r^{2z}}+\beta(z,L,
\phi_{i})\frac{d\vec{x}^2}{r^{2}}
+[\frac{(d-2)\beta(z,L,\phi_{i})-z\alpha(z,L,\phi_{i})}{d-2+z}]
\frac{dr^{2}}{r^{2}}.
\end{equation}
Then the Einstein equations reduce to
\begin{equation}
\alpha=\alpha(z,L,\phi_{i}),~~~~\beta=\beta(z,L,\phi_{i}).
\end{equation}
Once the values of $\phi_{i}$ are fixed by the $\phi$ equations of
motion, the above equations can be seen as two equations for $L$ and
$z$.

Thus we can conclude as follows: The Lifshitz symmetry of the
spacetime will be deformed (by changing the value of $z$) but not
broken once higher order corrections are incorporated. In
particular, if we can find a Lifshitz spacetime with certain $z$ and
$L$ for one action, then for any small deformations of the
parameters in the action we may find another solution with
$z^{\prime}, L^{\prime}$ which are nearby values of $z$ and $L$.
Conversely, variations of the action can only renormalize the
parameters $z$ and $L$ in the metric.
\subsection{Solutions in asymptotically Lifshitz spacetimes}\label{sec:2.2}
We will review the solutions which are asymptotic to Lifshitz metric
obtained in~\cite{Taylor:2008tg}, including both zero-temperature
and black brane cases. Consider the following action in
$(d+2)$-dimensional spacetime (without higher derivative
corrections)
\begin{equation}
\label{2eq9} S=\frac{1}{16\pi G_{d+2}}\int
d^{d+2}x\sqrt{-g}[R-2\Lambda-\frac{1}{2}\partial_{\mu}\phi\partial^{\mu}\phi
-\frac{1}{4}e^{\lambda\phi}F_{\mu\nu}F^{\mu\nu}],
\end{equation}
where $\Lambda$ is the cosmological constant and the matter fields
are a massless scalar and an abelian gauge field.

Such a theory admits the following zero-temperature solution whose
metric is Lifshitz-like
\begin{eqnarray}
\label{2eq13} &
&ds^{2}=L^{2}(-r^{2z}dt^{2}+\frac{dr^{2}}{r^{2}}+r^{2}\sum\limits^{d}_{i=1}dx^{2}_{i}),\nonumber\\
&
&F_{rt}=qe^{-\lambda\phi}r^{z-d-1},~~~e^{\lambda\phi}=r^{\lambda\sqrt{2(z-1)d}},\nonumber\\
& &\lambda^{2}=\frac{2d}{z-1},~~~q^{2}=2L^{2}(z-1)(z+d),\nonumber\\
& &\Lambda=-\frac{(z+d-1)(z+d)}{2L^{2}}
\end{eqnarray}
as well as black brane solution
\begin{equation}
ds^{2}=L^{2}(-r^{2z}f(r)dt^{2}+\frac{dr^{2}}{r^{2}f(r)}+r^{2}\sum
\limits^{d}_{i=1}dx^{2}_{i}),~~~
f(r)=1-\frac{r^{z+d}_{+}}{r^{z+d}},
\end{equation}
where the other fields in the black brane solution remain the same
as those in the zero-temperature solution. It is found that the AdS
spacetime is also a solution to the equations of motion with
$\phi=0$ and $F_{rt}=0$. Although the above metrics are
Lifshitz-like or asymptotically Lifshitz-like, such solutions cannot
be thought of as genuine gravity duals of Lifshitz fixed points, as
the dilaton is not constant. However, such solutions are of interest
themselves due to the black brane solution. We can study the
thermodynamic and hydrodynamic properties of such black branes,
which may be of help in understanding the genuine gravity duals of
Lifshitz fixed points at finite temperature.

In the last subsection, it was required that the gauge invariant
observables must be invariant under the full Lifshitz symmetry. Here
we can see that such a requirement can be released due to the
coupling between the dilaton and the gauge fields. The stress tensor
is
\begin{equation}
T_{\mu\nu}=-\frac{1}{2}g_{\mu\nu}(\frac{1}{2}\partial_{\rho}\phi\
\partial^{\rho}\phi+\frac{1}{4}e^{\lambda\phi}F_{\rho\sigma}F^{\rho\sigma})
+\frac{1}{2}\partial_{\mu}\phi\partial_{\nu}\phi+\frac{1}{2}e^{\lambda\phi}
F_{\mu\rho}{F_{\nu}}^{\rho}.
\end{equation}
The components of the stress tensor can be obtained by substituting
the values of the matter fields in~(\ref{2eq13})
\begin{equation}
T_{tt}=(z-1)(\frac{z}{2}+d)r^{2z},~~~T_{rr}=-\frac{z(z-1)}{2r^{2}},~~~T_{ii}
=\frac{1}{2}z(z-1)r^{2},
\end{equation}
which are all invariant under the Lifshitz symmetries. Once higher
derivative corrections are incorporated, the equations of motion of
the matter fields do not change and the additional part of the
stress tensor is comprised of the Riemann tensor of the background
geometry. Thus the stress tensor is still invariant under the
Lifshitz symmetries.
\section{$R^{2}$ corrections to zero-temperature backgrounds}
We shall study $R^{2}$ corrections to the zero-temperature Lifshitz
geometry in this section. Firstly we will obtain a solution in pure
Gauss-Bonnet gravity and then we will consider the action appearing
in~(\ref{2eq9}) plus Gauss-Bonnet corrections. It should be
emphasized that both of the solutions are exact, while we will
investigate $R^{2}$ corrected black brane solutions in the next
section by perturbative methods.

The general action containing the curvature squared corrections can
be written as
\begin{equation}
S=\frac{1}{16\pi G}\int
d^{D}x\sqrt{-g}[R-2\Lambda+L^{2}(\alpha_{1}R_{\mu\nu\rho\sigma}R^{\mu\nu\rho\sigma}
+\alpha_{2}R_{\mu\nu}R^{\mu\nu}+\alpha_{3}R^{2})]+S_{m},
\end{equation}
where $\alpha_{i}$ are arbitrary small coefficients and $S_{m}$
denotes the matter sector of the action. One specific
model--Gauss-Bonnet gravity--has provided many interesting results,
whose action is
\begin{equation}
\label{3eq2} S=\frac{1}{16\pi G}\int
d^{D}x\sqrt{-g}[R-2\Lambda+\frac{\lambda_{\rm
GB}}{2}L^{2}(R_{\mu\nu\rho\sigma}R^{\mu\nu\rho\sigma}
-4R_{\mu\nu}R^{\mu\nu}+R^{2})]+S_{m}.
\end{equation}
Several exact solutions of black holes in Gauss-Bonnet gravity have
been obtained, see e.g.~\cite{NO1, Cai:2001dz}. FRW-like solutions
and black holes for general five-dimensional $R^{2}$ gravity were
studied in~\cite{NO2}. The conformal anomaly from higher derivative
gravity in AdS/CFT correspondence was studied in~\cite{NO3}. From
now on we will focus on five-dimensional case as the Gauss-Bonnet
corrections are topological in four-dimensional spacetime and do not
play an important role. The Einstein equations derived
from~(\ref{3eq2}) are
\begin{equation}
\label{3eq3} R_{\mu\nu}-\frac{1}{2}Rg_{\mu\nu}=-\Lambda
g_{\mu\nu}+T^{\rm M}_{\mu\nu}+T^{\rm R}_{\mu\nu},
\end{equation}
where $T^{\rm M}_{\mu\nu}$ stands for the stress tensor of the
matter sector and $T^{R}_{\mu\nu}$ comes from the Gauss-Bonnet term
\begin{eqnarray}
\label{3eq4} T^{\rm R}_{\mu\nu}&=&\frac{\lambda_{\rm
GB}}{2}L^{2}[\frac{1}{2}g_{\mu\nu}(R_{\gamma\delta\lambda\sigma}R^{\gamma\delta\lambda\sigma}
-4R_{\gamma\delta}R^{\gamma\delta}+R^{2})-2RR_{\mu\nu}\nonumber\\&
&+4R_{\mu\gamma}{R^{\gamma}}_{\nu}+
4R^{\gamma\delta}R_{\gamma\mu\delta\nu}-2R_{\mu\gamma\delta\lambda}{R_{\nu}}^{\gamma\delta\lambda}].
\end{eqnarray}

\subsection{Solutions in pure Gauss-Bonnet gravity}
First let us consider Lifshitz-like solutions in pure Gauss-Bonnet
gravity, i.e.without introducing matter fields. The Einstein
equations~(\ref{3eq3}) turn out to be
\begin{equation}
R_{\mu\nu}-\frac{1}{2}Rg_{\mu\nu}=-\Lambda g_{\mu\nu}+T^{\rm
R}_{\mu\nu},
\end{equation}
where $T^{\rm R}_{\mu\nu}$ has been given in~(\ref{3eq4}). The
Lifshitz background can be written as
\begin{equation}
ds^{2}=L^{2}[-\frac{dt^{2}}{r^{2z}}+\frac{1}{r^{2}}(dr^{2}+dx^{2}_{1}+dx^{2}_{2}+dx^{2}_{3})].
\end{equation}
The non-vanishing components of the Ricci tensor are
\begin{equation}
R_{tt}=\frac{z(z+3)}{r^{2z}},~~~R_{rr}=-\frac{z^{2}+3}{r^2},~~~R_{ii}=-\frac{z+3}{r^{2}},~~~i=1,2,3,
\end{equation}
and the non-vanishing components of the stress tensor
\begin{equation}
T^{\rm R}_{tt}=-\frac{6\lambda_{\rm GB}}{r^{2}},~~~T^{\rm
R}_{rr}=\frac{6\lambda_{\rm GB}z}{r^{2}},~~~T^{\rm
R}_{ii}=\frac{2\lambda_{\rm GB}z(z+2)}{r^2}.
\end{equation}

Thus we can easily find that the Lifshitz background is a solution
to the Einstein equations when the Gauss-Bonnet coupling constant
$\lambda_{\rm GB}$ and the cosmological constant $\Lambda$ take the
following values
\begin{equation}
\lambda_{\rm GB}=\frac{1}{2},~~~\Lambda=-\frac{3}{L^{2}}.
\end{equation}
Here are some remarks on this solution:
\begin{itemize}
\item When z=1, the metric reduces to AdS, which is a solution to
the Einstein equations for any values of $\lambda_{\rm GB}$ and
$\Lambda$.
\item In the literatures studying Gauss-Bonnet black holes in
AdS and dS spacetimes~\cite{Cai:2001dz}, in order to obtain a
meaningful black hole solution, the Gauss-Bonnet coupling constant
$\lambda_{\rm GB}$ should have an upper bound $\lambda^{\rm
upper}_{\rm GB}=1/4$. Here $\lambda_{\rm GB}$ goes beyond this
bound. We will go back to this issue when discussing the ratio of
shear viscosity to entropy density.
\item In~\cite{Adams:2008zk}, an exact solution of Lifshitz geometry
was obtained in pure $R^{2}$ gravity, with the coefficient in front
of the $R^{2}$ term and the cosmological constant
$$c_{1}=\frac{L^{2}}{2z^{2}+4z+6},~~~\Lambda=\frac{z^{2}+2z+3}{L^{2}}.$$
We can see that when the space is large, so is the higher order
corrections, and vice versa. This invalidates the perturbative
description, as these solutions balance the curvature terms against
curvature squared terms, the quadratic approximation cannot be
expected to be reliable. Here we have similar situations and just
as~\cite{Adams:2008zk}, we can also expect that a non-trivial matter
sector may solve this problem.
\item In~\cite{NO1, Cai:2001dz}, exact solutions of black holes in pure
Gauss-Bonnet gravity were obtained. However, due to the Birkhoff
theorem, we cannot find exact solutions of black holes with
anisotropic scaling symmetry in pure Gauss-Bonnet gravity.
\end{itemize}
\subsection{Solutions with non-trivial matter fields}
We will consider Lifshitz-like solutions with non-trivial matter
fields in this subsection. To be concrete, we shall add Gauss-Bonnet
corrections to the effective action~(\ref{2eq9}) and try to find the
corresponding solutions. Note that here the solutions are exact
while we will study the corrections in the black brane case by
perturbative methods in the next section.

Now the Einstein equations are given by~(\ref{3eq3})
$$R_{\mu\nu}-\frac{1}{2}Rg_{\mu\nu}=-\Lambda g_{\mu\nu}+T^{\rm
M}_{\mu\nu}+T^{\rm R}_{\mu\nu},$$ where the stress tensor of the
matter sector is
$$T^{\rm M}_{\mu\nu}=-\frac{1}{2}g_{\mu\nu}(\frac{1}{2}\partial_{\rho}\phi\
\partial^{\rho}\phi+\frac{1}{4}e^{\lambda\phi}F_{\rho\sigma}F^{\rho\sigma})
+\frac{1}{2}\partial_{\mu}\phi\partial_{\nu}\phi+\frac{1}{2}e^{\lambda\phi}
F_{\mu\rho}{F_{\nu}}^{\rho}$$ and $T^{\rm R}_{\mu\nu}$ has been
shown in~(\ref{3eq4}). In addition, the equations of motion for the
matter fields are
\begin{equation}
\partial_{\mu}(\sqrt{-g}e^{\lambda\phi}F^{\mu\nu})=0,
\end{equation}
\begin{equation}
\partial_{\mu}(\sqrt{-g}\partial^{\mu}\phi)-\frac{\lambda}{4}
\sqrt{-g}e^{\lambda\phi}F_{\mu\nu}F^{\mu\nu}=0.
\end{equation}
Here we still make the ansatz for the metric
$$ds^{2}=L^{2}[-\frac{dt^{2}}{r^{2z}}+\frac{1}{r^{2}}
(dr^{2}+dx^{2}_{1}+dx^{2}_{2}+dx^{2}_{3})].$$

After solving the equations of motion, we can find the following
solution
\begin{eqnarray}
& &\phi=\pm\sqrt{6(1-2\lambda_{\rm GB})(z-1)}\log r,~~~~F_{rt}=
qe^{-\lambda\phi}r^{2-z},\nonumber\\
& &\lambda^{2}=\frac{6}{(1-2\lambda_{\rm
GB})(z-1)},~~~~q^{2}=2(1-2\lambda_{\rm
GB})(z-1)(z+3)L^{2},\nonumber\\
& &\Lambda=-\frac{(z+2)(z+3)}{2L^{2}}+\frac{\lambda_{\rm
GB}z(z+5)}{L^{2}}.
\end{eqnarray}
Note that here the dynamical exponent $z$ has been renormalized.
Furthermore, $\lambda_{\rm GB}$ should have an upper bound $1/2$ to
ensure a physical solution. Let us denote $z=z_{0}+\delta z$ where
$z_{0}$ is the dynamical exponent in the Einstein-matter theory. By
fixing the cosmological constant $\Lambda$ and solving the
linearized equations, we can arrive at
\begin{equation}
\delta z=\frac{2\lambda_{\rm GB}z_{0}(z_{0}+5)}{(1-2\lambda_{\rm
GB})(2z_{0}+5)}.
\end{equation}

The $z_{0}=1$ case should be treated separately. As can be seen from
the unperturbed solution~(\ref{2eq13}), when $z_{0}=1$, both the
dilaton and the gauge field strength vanish, then the theory reduces
to pure Einstein gravity. The non-renormalization of the AdS case is
a well known fact. Furthermore, in~\cite{Adams:2008zk}, when
considering four-dimensional Lifshitz spacetime, it was found that
the $z_{0}=2$ case seemed to be protected. But there was no sign of
an extra symmetry protecting this solution and it could be
renormalized under certain ad-hoc higher order terms. Here we can
see that $z$ can also be renormalized in $z_{0}=2$ case, so
$z_{0}=2$ is nothing special compared to other cases, which supports
their argument.
\section{$R^{2}$ corrections to black branes}
In this section we consider Gauss-Bonnet corrections to black
branes. Unfortunately, it is quite difficult to find an exact
solution which is asymptotic to Lifshitz spacetime in Gauss-Bonnet
gravity. Then we have to solve the equations of motion
perturbatively, following~\cite{Kats:2007mq}.

Considering the following action
\begin{eqnarray}
\label{4eq1} S&=&\frac{1}{16\pi G_{5}}\int
d^{5}x\sqrt{-g}[R-2\Lambda-\frac{1}{2}\partial_{\mu}\phi\partial^{\mu}\phi
-\frac{1}{4}e^{\lambda\phi}F_{\mu\nu}F^{\mu\nu}\nonumber\\
& &+\frac{\lambda_{\rm
GB}}{2}L^{2}(R_{\mu\nu\rho\sigma}R^{\mu\nu\rho\sigma}
-4R_{\mu\nu}R^{\mu\nu}+R^{2})],
\end{eqnarray}
the corresponding equations of motion remain the same as before
\begin{eqnarray}
& &R_{\mu\nu}-\frac{1}{2}Rg_{\mu\nu}=-\Lambda g_{\mu\nu}+T^{\rm
M}_{\mu\nu}+T^{\rm R}_{\mu\nu},\nonumber\\
& &T^{\rm
M}_{\mu\nu}=-\frac{1}{2}g_{\mu\nu}(\frac{1}{2}\partial_{\rho}\phi\
\partial^{\rho}\phi+\frac{1}{4}e^{\lambda\phi}F_{\rho\sigma}F^{\rho\sigma})
+\frac{1}{2}\partial_{\mu}\phi\partial_{\nu}\phi+\frac{1}{2}e^{\lambda\phi}
F_{\mu\rho}{F_{\nu}}^{\rho},\nonumber\\
& &T^{\rm R}_{\mu\nu}=\frac{\lambda_{\rm
GB}}{2}L^{2}[\frac{1}{2}g_{\mu\nu}(R_{\gamma\delta\lambda\sigma}R^{\gamma\delta\lambda\sigma}
-4R_{\gamma\delta}R^{\gamma\delta}+R^{2})-2RR_{\mu\nu}\nonumber\\&
&+4R_{\mu\gamma}{R^{\gamma}}_{\nu}+
4R^{\gamma\delta}R_{\gamma\mu\delta\nu}-
2R_{\mu\gamma\delta\lambda}{R_{\nu}}^{\gamma\delta\lambda}],\nonumber\\
& &\partial_{\mu}(\sqrt{-g}e^{\lambda\phi}F^{\mu\nu})=0,\nonumber\\
& &\partial_{\mu}(\sqrt{-g}\partial^{\mu}\phi)-\frac{\lambda}{4}
\sqrt{-g}e^{\lambda\phi}F_{\mu\nu}F^{\mu\nu}=0.
\end{eqnarray}
Note that the equations of motion for the matter fields do not
change after incorporating the Gauss-Bonnet corrections.

Let us focus on the right hand side of the Einstein equations. Since
we are trying to solve the equations at the leading order of
$\lambda_{\rm GB}$, we can substitute the unperturbed metric
$$ds^{2}=L^{2}[-\frac{f(r)}{r^{2z}}dt^{2}+\frac{dr^{2}}{r^{2}f(r)}
+\frac{1}{r^{2}}(dx^{2}_{1}+dx^{2}_{2}+dx^{2}_{3})],~~~
f(r)=1-\frac{r^{z+3}}{r^{z+3}_{+}}$$ into $T^{R}_{\mu\nu}$.
Furthermore, we neglect the backreactions of the Gauss-Bonnet
corrections to the matter fields and substitute the unperturbed
values of those fields into $T^{\rm M}_{\mu\nu}$. The ansatz for the
metric is
\begin{equation}
ds^{2}=\frac{L^{2}}{r^{2}}[-e^{2a(r)}dt^{2}+e^{-2b(r)}dr^{2}+dx^{2}_{1}
+dx^{2}_{2}+dx^{2}_{3}].
\end{equation}
The components of the Ricci tensor can be combined as
\begin{equation}
R^{t}_{t}-R^{r}_{r}=\frac{3}{L^{2}}e^{2b(r)}r(a^{\prime}(r)-b^{\prime}(r)),
\end{equation}
\begin{equation}
\frac{1}{3}(R^{t}_{t}-R^{r}_{r})-R^{x}_{x}=-\frac{1}{L^{2}}
(\frac{e^{2b(r)}}{r^{4}})^{\prime}r^{5},
\end{equation}
where the prime stands for derivative with respect to $r$. On the
other hand, we have the following expressions due to the Einstein
equations
\begin{equation}
R^{t}_{t}-R^{r}_{r}={T^{\rm M}}^{t}_{t}+{T^{\rm R}}^{t}_{t}-{T^{\rm
M}}^{r}_{r}-{T^{\rm R}}^{r}_{r},
\end{equation}
\begin{equation}
\frac{1}{3}(R^{t}_{t}-R^{r}_{r})-R^{x}_{x}=\frac{2}{3}({T^{\rm
M}}^{t}_{t} +{T^{\rm M}}^{r}_{r}-\Lambda).
\end{equation}
Therefore,
\begin{equation}
e^{2b(r)}=-\frac{2}{3}L^{2}r^{4}[\int\frac{dr}{r^{5}}({T^{\rm
M}}^{t}_{t} +{T^{\rm M}}^{r}_{r}-\Lambda)+\rm const],
\end{equation}
\begin{equation}
a(r)=b(r)+\frac{L^{2}}{3}\int\frac{dr}{r}e^{-2b(r)}({T^{\rm
M}}^{t}_{t}+{T^{\rm R}}^{t}_{t}-{T^{\rm M}}^{r}_{r}-{T^{\rm
R}}^{r}_{r}).
\end{equation}

After substituting the unperturbed metric and matter fields, we can
obtain the following perturbative black brane solution
\begin{equation}
\label{4eq10}
ds^{2}=L^{2}[-\frac{f(r)}{r^{2z}}h(r)dt^{2}+\frac{dr^{2}}{r^{2}f(r)}+
\frac{1}{r^{2}}(dx^{2}_{1}+dx^{2}_{2}+dx^{2}_{3})],
\end{equation}
where
\begin{eqnarray}
& &z=z_{0}+2\lambda_{\rm
GB}(z_{0}-1),~~~h(r)=\exp[4\lambda_{\rm GB}\frac{z_{0}-1}{z_{0}+3}(\frac{r}{r_{+}})^{z_{0}+3}],\nonumber\\
& &f(r)=1-(\frac{r}{r_{+}})^{z_{0}+3}+\lambda_{\rm
GB}[1-(\frac{r}{r_{+}})^{z_{0}+3}]^{2}.
\end{eqnarray}
One can check that this solution agrees with the one appearing
in~\cite{Kats:2007mq} when $z_{0}=1$. For general cases, the horizon
still locates at $r=r_{+}$ and the Hawking temperature is
\begin{equation}
T_{H}=\frac{1}{4\pi}\frac{z_{0}+3}{r^{z}_{+}}(1+2\lambda_{\rm
GB}\frac{z_{0}-1}{z_{0}+3}).
\end{equation}
For black branes in Gauss-Bonnet gravity, the area law for entropy
still holds~\cite{Jacobson:1993xs}, then
\begin{equation}
S_{BH}=\frac{1}{4\pi}\frac{L^{3}V_{3}}{r^{3}_{+}},
\end{equation}
where $V_{3}$ denotes the volume of the spatial directions.

\section{Calculating $\eta/s$}
The AdS/CFT correspondence has provided us an efficient way to study
the dynamics of strongly coupled gauge theories. One remarkable
example is the calculation of the ratio of the shear viscosity over
entropy density $\eta/s$. It has been found that
$$\frac{\eta}{s}=\frac{1}{4\pi}$$
is a universal result for all gauge theories with Einstein gravity
duals in the large N limit. Furthermore, it was conjectured that
$1/4\pi$ is a universal lower bound for all materials, which is
known as the KSS bound~\cite{Kovtun:2004de}. Later the authors
of~\cite{Kats:2007mq, Brigante:2007nu, Brigante:2008gz} calculated
the ratio in $R^{2}$ gravity and found that the lower bound was
violated. A new lower bound $4/25\pi$ was proposed
in~\cite{Brigante:2008gz} by considering the causality of the dual
field theory. For more discussions on violation of the KSS bound in
higher derivative gravity, see~\cite{more}.

It was conjectured that the shear viscosity is completely determined
by the effective coupling of the transverse gravitons on the horizon
in the dual gravity description~\cite{Brustein:2007jj}. This was
confirmed in~\cite{Iqbal:2008by} via the scalar membrane paradigm
and in~\cite{Cai:2008ph} by calculating the on-shell action of the
transverse gravitons. Such an effective action in a given background
was assumed to be a minimally coupled massless scalar with an
effective coupling which depends on the radial coordinate, while in
Einstein gravity the effective coupling is a constant. However, this
formalism is not covariant under coordinate transformations, then
the coordinate system of the background geometry also affects the
form of the action of transverse gravitons. In~\cite{Cai:2009zv}, a
new formalism was proposed, where a new three-dimensional effective
metric $\tilde{g}_{\mu\nu}$ was introduced and the transverse
gravitons were minimally coupled to this new effective metric. The
action in this new formalism can take a covariant form. Similar
discussions on this issue were also presented
in~\cite{Banerjee:2009wg}.

We shall calculate the shear viscosity of field theory with $R^{2}$
corrected black brane dual using the approach proposed
in~\cite{Cai:2009zv}. As a non-relativistic theory, the causality of
the boundary field theory cannot be treated as a constraint on the
lower bound of $\eta/s$.
\subsection{Shear viscosity from the effective coupling of transverse gravitons}
The shear viscosity can be calculated via the Kubo formula
\begin{equation}
\eta=\lim\limits_{\omega\rightarrow0}\frac{1}{2\omega
i}(G^{A}_{x_{1}x_{2},x_{1}x_{2}}(\omega,0)-G^{R}
_{x_{1}x_{2},x_{1}x_{2}}(\omega,0)),
\end{equation}
where the retarded Green's function $G^{R}_{\mu\nu,\lambda\rho}$ is
defined by
\begin{equation}
G^{R}_{\mu\nu,\lambda\rho}=-i\int d^{4}xe^{-ik\cdot x}\theta(t)
<[T_{\mu\nu}(x),T_{\lambda\rho}(0)]>,
\end{equation}
and the advanced Green's function satisfies
$G^{A}_{\mu\nu,\lambda\rho}(k)={G^{R}_{\mu\nu,\lambda\rho}(k)}^{\ast}$.
These Green's functions are defined on the field theory side.
According to the field-operator correspondence, such Green's
functions can be calculated through the effective action of the
gravitons of the dual gravity theory.

We can choose spatial coordinates so that the momentum of the
perturbation points along the $x_{3}\equiv z$ axis. Considering
tensor perturbation $h_{12}=h_{12}(t,u,z)$ with $u$ being the radial
coordinate, we denote $\phi=h^{1}_{2}$ and write $\phi$ as
$\phi(t,u,z)=\phi(u)e^{-i\omega t+ipz}$. For gravity theories in
which the transverse gravitons can be decoupled from other
perturbations, the effective bulk action of the transverse gravitons
can be written in a general form
\begin{equation}
S=\frac{V_{1,2}}{16\pi G}(-\frac{1}{2})\int
d^{3}x\sqrt{-\tilde{g}}(\tilde{K}(u)\tilde{g}^{MN}
\tilde{\nabla}_{M}\phi\tilde{\nabla}_{N}\phi+m^{2}\phi^{2})
\end{equation}
up to some total derivatives. Here $\tilde{g}_{MN}$ is a
three-dimensional effective metric, $m$ is an effective mass and
$\tilde{\nabla}_{M}$ is the covariant derivative using
$\tilde{g}_{MN}$. Notice that $\phi$ is a scalar in the three
dimensions $t,u,z$, while it is not a scalar in the whole five
dimensions. We write the action in the three-dimensional form so
that it is general covariant and $\tilde{K}(u)$ is a scalar under
general coordinate transformations. It should be pointed out that
this is not the ordinary dimensional reduction. In the following we
will use $g_{\mu\nu}$ to denote the whole five-dimensional
background.

Recalling the corrected black brane metric in~(\ref{4eq10}) and
performing the following coordinate transformations
\begin{equation}
\rho=\frac{1}{r},~~~\rho_{+}=\frac{1}{r_{+}},~~~
(\frac{\rho_{+}}{\rho})^{z_{0}+3}=u^{2},
\end{equation}
the black brane metric metric turns out to be
\begin{equation}
\label{5eq5}
ds^{2}=L^{2}[-g(u)(1-u)dt^{2}+\frac{1}{h(u)(1-u)}du^{2}+\frac{\rho_{+}^{2}}{u^{A}}
(dx^{2}_{1}+dx^{2}_{2}+dx^{2}_{3})],
\end{equation}
where
\begin{eqnarray}
& &g(u)=\rho^{2z}_{+}u^{-\frac{4z}{z_{0}+3}}(1+u)(1+\lambda_{\rm
GB}(1-u^{2}))\exp[4\lambda_{\rm
GB}\frac{z_{0}-1}{z_{0}+3}u^{2}],\nonumber\\
& &h(u)=\frac{1}{4}(z_{0}+3)^{2}u^{2}(1+u)(1+\lambda_{\rm
GB}(1-u^{2})),~~~A=\frac{4}{z_{0}+3}.
\end{eqnarray}
Then the horizon of the black brane locates at $u=1$ and the
boundary locates at $u=0$.

Following~\cite{Cai:2009zv}, we write down the action of the
transverse gravitons in momentum space
\begin{eqnarray}
\label{5eq7} S&=&\frac{V_{1,2}}{16\pi
G}(-\frac{1}{2})\int\frac{dwdp}{(2\pi)^{2}}du\sqrt{-\tilde{g}}[
\tilde{K}(u)(\tilde{g}^{uu}\phi^{\prime}\phi^{\prime}\nonumber\\
&
&+w^{2}\tilde{g}^{tt}\phi^{2}+p^{2}\tilde{g}^{zz}\phi^{2})+m^{2}\phi^{2}],
\end{eqnarray}
where
\begin{eqnarray}
&
&\phi(t,u,z)=\int\frac{dwdp}{(2\pi)^{2}}\phi(u;k)e^{-iwt+ipz},\nonumber\\
& &k=(w,0,0,p),~~~\phi(u;-k)=\phi^{\ast}(u;k),
\end{eqnarray}
and the prime denotes derivative with respect to $u$. The
corresponding equation of motion is given by
\begin{equation}
\phi^{\prime\prime}(u;k)+A(u)\phi^{\prime}(u;k)+B(u)\phi(u;k)=0,
\end{equation}
where
\begin{equation}
A(u)=\frac{{(\sqrt{-\tilde{g}}\tilde{K}(u)\tilde{g}^{uu})}^{\prime}}
{\sqrt{-\tilde{g}}\tilde{K}(u)\tilde{g}^{uu}},~~~B(u)=-\tilde{g}_{uu}(\tilde{g}^{tt}w^{2}
+\tilde{g}^{zz}p^{2}+\frac{m^{2}}{\tilde{K}(u)}).
\end{equation}
Furthermore, by repeating the calculations in Section II
of~\cite{Cai:2009zv}, we can find that here we still have the
following formula for $\eta$
\begin{equation}
\eta=\frac{1}{16\pi G}(\sqrt{\tilde{g_{zz}}}\tilde{K}(u))|_{u=1}.
\end{equation}

Next we shall calculate the effective action of the transverse
gravitons on the background~(\ref{5eq5}). From the first order
Einstein equations we can see that the transverse gravitons can get
decoupled from other perturbations. Then we can obtain the effective
action of the transverse gravitons by keeping quadratic terms of
$\phi$ in the original action~(\ref{4eq1}). The action can be
written in the form of~(\ref{5eq7}) with three-dimensional effective
metric
\begin{equation}
\tilde{g}^{uu}=(1+\frac{\lambda_{\rm
GB}}{2}\frac{Ag^{\prime}_{tt}g^{uu}}{ug_{tt}})g^{uu},
\end{equation}
\begin{equation}
\label{5eq13} \tilde{g}^{tt}=[1+\frac{\lambda_{\rm
GB}}{2}(\frac{Ag^{\prime
uu}}{u}-\frac{(A^{2}+2A)g^{uu}}{u^{2}})]g^{tt},
\end{equation}
\begin{equation}
\label{5eq14} \tilde{g}^{zz}=[1+\frac{\lambda_{\rm
GB}}{2}(\frac{g^{\prime2}_{tt}g^{uu}}{g^{2}_{tt}}
-\frac{g^{\prime}_{tt}g^{\prime
uu}}{g_{tt}}-\frac{2g^{uu}g^{\prime\prime}_{tt}}{g_{tt}})]g^{zz}.
\end{equation}
The $m^{2}$ term vanishes due to the Einstein equations of the
background metric. Note that when $z_{0}=1$, i.e. $A=1$, the above
expressions agree with those obtained in~\cite{Cai:2009zv} with
vanishing dilaton field.

By using the formula $\tilde{K}(u)=\sqrt{-g}/\sqrt{-\tilde{g}}$ and
recalling the fact that the area law still holds for black branes in
Gauss-Bonnet gravity, we can arrive at the final result
\begin{equation}
\frac{\eta}{s}=\frac{1}{4\pi}[1-\frac{\lambda_{\rm
GB}}{2}Ah(1)]=\frac{1}{4\pi}[1-(z_{0}+3)\lambda_{\rm GB}].
\end{equation}
Here are some remarks on this result:
\begin{itemize}
\item When $z_{0}=1$, the black brane metric is asymptotically AdS.
We can recover the result obtained in~\cite{Brigante:2007nu}.
\item For general $z_{0}\neq1$, in order to obtain a non-vanishing
$\eta/s$, $\lambda_{\rm GB}$ should have an upper bound
$1/(z_{0}+3)$. The upper bound of $\lambda_{\rm GB}$ was discussed
in~\cite{Ge:2009eh} where it was found to be $1/4$ by the
constraints of causality and stability. Here a similar upper bound
in non-relativistic theory requires further understanding.
\item In the literatures discussing the ratio of shear viscosity
over entropy density in higher derivative theory of gravity, the new
lower bound of $\eta/s$--$4/25\pi$--can be obtained by considering
the causality of the boundary field theory. However, here we cannot
take such a constraint as the dual field theory is non-relativistic.
We will address it in detail in next subsection. \end{itemize}
\subsection{Causality cannot be a constraint}
It was discovered that the KSS bound can be violated in $R^{2}$
gravity, but the causality of the boundary field theory can
constrain the parameters and introduce a new lower
bound~\cite{Brigante:2007nu, Brigante:2008gz}. But here we will see
that causality cannot be a constraint in a non-relativistic theory.

Following~\cite{Cai:2009zv}, we can transform the action of the
transverse gravitons into a minimally coupled form
\begin{equation}
S=\frac{V_{1,2}}{16\pi G}{-\frac{1}{2}}\int
d^{3}x\sqrt{-\bar{g}}(\bar{g}^{MN}\partial_{M}\phi\partial_{N}\phi+\bar{m}^{2}\phi^{2}),
\end{equation}
with
\begin{equation}
\bar{g}^{MN}=\tilde{K}(u)^{-2}\tilde{g}^{MN},~~~\bar{m}^{2}=\tilde{K}(u)^{-3}m^{2}.
\end{equation}
For the case at hand, $\bar{m}^{2}$ term vanishes, so the equation
of motion is
\begin{equation}
\bar{g}^{MN}\bar{\nabla}_{M}\bar{\nabla}_{N}\phi=0.
\end{equation}
Then we can apply the geometrical optics approximation in the large
momentum limit. The wave function is written in the form
$\phi=\phi_{en}(t,u,z)e^{i\theta(t,u,z)}$, where $\phi_{en}$ stands
for a slowly changing envelope function and $\theta$ is a rapidly
varying phase function. Expanding the equation of motion at leading
order, we obtain
\begin{equation}
\label{5eq19} \frac{dx^{M}}{ds}\frac{dx^{N}}{ds}\bar{g}_{MN}=0,
\end{equation}
where $dx^{M}/ds=\bar{g}^{MN}\bar{\nabla}_{N}\theta$.

Due to translation symmetries in $t$ and $z$ directions,
$\omega=i\bar{\nabla}_{t}\theta$ and $q=-i\bar{\nabla}_{z}\theta$
are still conserved integrals of motion along the geodesic.
Then~(\ref{5eq19}) can be written as
\begin{equation}
(\frac{du}{ds})^{2}=(-\bar{g}^{tt}\bar{g}^{uu}q^{2})[\frac{\omega^{2}}{q^{2}}
-\frac{\bar{g}^{zz}}{-\bar{g}^{tt}}].
\end{equation}
If we assume $q^{2}>0$ and denote
$\tilde{s}=s\sqrt{-\bar{g}^{tt}\bar{g}^{uu}q^{2}}$, we can get
\begin{equation}
(\frac{du}{d\tilde{s}})^{2}=\frac{\omega^{2}}{q^{2}}
-\frac{\bar{g}^{zz}}{-\bar{g}^{tt}}.
\end{equation}
This equation describes a one-dimensional system with a particle of
energy $\frac{\omega^{2}}{q^{2}}$ moving in a potential
$\frac{\bar{g}^{zz}}{-\bar{g}^{tt}}$. The effective geometry can be
expressed as
\begin{equation}
ds^{2}=\bar{g}_{MN}dx^{M}dx^{N}=-\bar{g}_{tt}(-dt^{2}+\frac{1}{c^{2}_{g}}dz^{2})+
\bar{g}_{uu}du^{2}, \end{equation} where
\begin{equation}
c^{2}_{g}=\frac{\bar{g}^{zz}}{-\bar{g}^{tt}}=\frac{\tilde{g}^{zz}}{-\tilde{g}^{tt}}
\end{equation}
denotes the local ``speed of graviton'' on constant $u$
hypersurface.

In the relativistic cases, $c^{2}_{g}=0$ on the horizon and
$c^{2}_{g}=1$ at infinity. One can expand $c^{2}_{g}$ near the
boundary and require that it should be smaller than one to avoid
causality violation. This requirement gives a new lower bound on
$\eta/s$. But here the situation is quite different, as it can be
easily seen that $c^{2}_{g}\rightarrow\infty$ at the boundary $u=0$
by using~(\ref{5eq13}) and~(\ref{5eq14}). This should not be
surprising as the boundary field theory is non-relativistic. So we
cannot take causality as a constraint on $\eta/s$.
\section{Summary and discussion}
We study $R^{2}$ corrections to asymptotically Lifshitz spacetimes
in five dimensions. For the zero-temperature background, we obtain
exact solutions both in pure Gauss-Bonnet gravity and in
Gauss-Bonnet gravity with non-trivial matter. In the latter case we
find that the dynamical exponent $z$ undergoes a finite
renormalization. For the finite-temperature background, we obtain
perturbative solutions in Gauss-Bonnet theory with non-trivial
matter. The ratio of shear viscosity over entropy density is also
calculated. It violates the KSS bound but here causality cannot be
treated as a constraint due to the non-relativistic nature of the
boundary theory.

The violation of the KSS bound in non-relativistic theory with
higher derivative correction was already observed
in~\cite{Adams:2008zk}, where they obtained
\begin{equation}
\frac{\eta}{s}=\frac{1}{4\pi}(1-\frac{1}{2N})
\end{equation}
for asymptotically Schr\"{o}dinger black holes. Here $N$ denotes the
rank of the gauge group and this result is the same as the
relativistic counterparts. Since the near horizon geometry of the
Schr\"{o}dinger black holes is the same as that of usual AdS black
holes, one can safely obtain the above result following the
arguments in ~\cite{Iqbal:2008by}. However, the result obtained in
this paper is different from the relativistic case. According
to~\cite{Iqbal:2008by}, this is natural as the Lifshitz black branes
and the AdS counterparts have different near horizon structures.
Furthermore, due to the difficulty of embedding the Lifshitz
background into string/M theory~\cite{Li:2009pf}, the corrections to
$\eta/s$ for Lifshitz black branes are difficult to evaluate in the
context of string /M theory.

Recently there have been several interesting discussions on higher
derivative corrections to $\eta/s$, see e.g.~\cite{Buchel:2008vz,
Cremonini:2009sy, Buchel:2009tt, Hofman:2009ug}. The causality of
the boundary field theory plays an important role in constraining
$\eta/s$. Here causality cannot be constraint since the boundary
field theory is non-relativistic. However, as argued
in~\cite{Adams:2008zk}, the observations in~\cite{Adams:2006sv}
would suggest that the problem with violations of KSS bound are as
much about unitarity and locality as about causality, and should
persist in the non-relativistic limit. The unitarity and locality
might be served as new constraints on the KSS bound in
non-relativistic theory and it would be interesting to investigate
this topic in future.

\bigskip \goodbreak \centerline{\bf Acknowledgements}
\noindent We thank Rong-Gen Cai and Ya-Wen Sun for valuable
discussions. This work is supported by the Korea Science and
Engineering Foundation(KOSEF) grant funded by the Korea
government(MEST) through the Center for Quantum Spacetime(CQUeST) of
Sogang University with grant number R11-2005-021.



\end{document}